# Graphene, a material with exceptional electronic properties

Modou B. NDIAYE, Anthony GEHA, Yago AGUADO

Modou_Bousso.Ndiaye@etu.sorbonne-universite.fr
anthony.gerges_geha@etu.sorbonne-universite.fr
yago.aguado@etu.sorbonne-universite.fr

**Abstract**

In this document we explore graphene, a two-dimensional material with remarkable properties. We center our discussion around its electronic characteristics and their applications. We begin by giving a simple electronic model which will then allow us to discuss its conductivity, electronic mobility and the peculiar situation we find at the Fermi level. We then move to the limitations imposed by the electronic structure, notably, the absence of a gap in the band. Finally, we take a look at a very active research area: modifications of the band structure of graphene and its potential applications.

## 1 Introduction

First isolated in 2004 by a team at the University of Manchester (Geim and Novoselov), graphene is a material with many surprising properties. After Geim and Noboselov's work, researchers have been able to show more and more exceptional characteristics and applications: semiconductivity; a great mechanical resistance; thermal conductivity; the quantum Hall effect; biomaterials; superconductivity for rotated bilayer graphene... One of the most interesting aspects is its electronic structure and the properties it generates, which is what this document addresses. Before delving into the subject, let us briefly present some of the defining traits of graphene.

Graphene is a two-dimensional, i.e. a single layer of atoms, honeycomb semi-metal composed uniquely of carbon and held together by covalent bonds 1.42 Å long. Structurally, it is organised in a hexagonal lattice, in which each carbon atom is bonded to its three nearest neighbours. Generally, 2-dimensional lattices are insulators, but graphene is an exception, it is semi-metallic. Putting together layers of graphene we get graphite, a very common material found, for example, in pencils, and which is a good conductor of heat and electricity, but whose potentialn in resarch and industry is very far form that of graphene. Indeed, graphene has shown remarkable properties in a number of domains, and many believe it will bring a revolution in electronics, through the creation of graphene supercapacitors; superconductivity, by exploiting the bilayer superconductivity that was shown in 2018 and chemical research, thanks to its high reactivity, to mention just a few active research areas.

This bibliographic project centers around the electronic properties of graphene. We first describe and propose a simple model of the electronic structure, as well as discuss its crystallographic structure and the electronic configuration that emerges. This will allow us to present with a proper background the advantages of graphene. Then, we will address a shortcoming, which is the absence of band gap, and discuss a way to "correct it".

Finally, because of their interest and relevance in the current research landscape, we briefly present the spin-orbit coupling and spin-Hall effect as well as a discussion of the importance of the existence of imperfections and disorder in graphene.

## 2 Electronic Properties of Graphene

### 2.1 Crystallographic Structure of graphene

Graphene exists as a single atomic layer composed exclusively of carbon atoms. The associated Bravais lattice takes the form of a triangular arrangement, characterized by a recurring pattern of two distinct atoms, A and B, present in each elementary unit cell. This unique crystalline organization gives rise to a honeycomb structure, emblematic of graphene. (Figure 1).

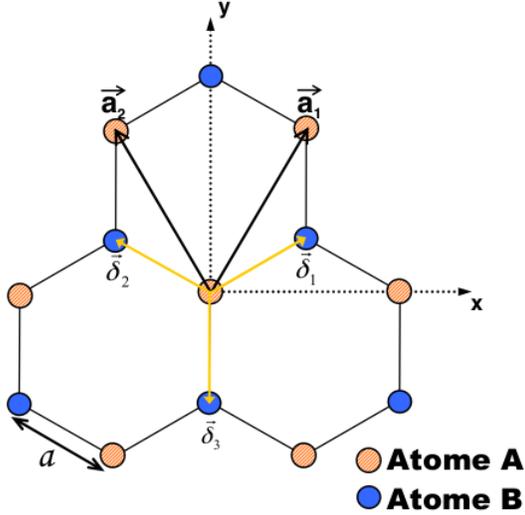
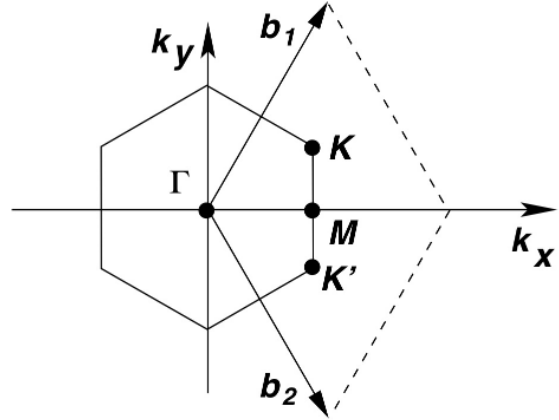

Figure 1: The honeycomb crystal structure of monolayer graphene. Atoms A and B do not have the same environment: B atoms have their first neighbors arranged in a triangle with the base at the bottom and pointing upwards, while A-type atoms have their first neighbors arranged in a triangle with the base at the top and pointing downwards.

We define the primitive lattice vectors vectors by :

$$\mathbf{a_1} = \frac{a}{2}(\sqrt{3}, 3), \quad \mathbf{a_2} = \frac{a}{2}(-\sqrt{3}, 3) \quad (2.1)$$

Where $a \approx 1.42$ Å is the distance between two neighboring atoms. We also define the vectors connecting an A atom with its three nearest neighbors (B atoms) :

$$\delta_1 = \frac{a}{2}(\sqrt{3}, 1), \quad \delta_2 = \frac{a}{2}(-\sqrt{3}, 1), \quad \delta_3 = a(0, -1) \quad (2.2)$$

Now, we express the vectors of the reciprocal lattice. Primitive reciprocal lattice vectors b1 and b2 satisfying

$$\mathbf{a_1}\mathbf{b_1} = \mathbf{a_2}\mathbf{b_2} = 2\pi \quad (2.3)$$

and

$$\mathbf{a_1}\mathbf{b_2} = \mathbf{a_2}\mathbf{b_1} = 0 \quad (2.4)$$

are given by :

$$\mathbf{b_1} = \frac{2\pi}{3a}(\sqrt{3}, 1), \quad \mathbf{b_2} = \frac{2\pi}{3a}(-\sqrt{3}, 1) \quad (2.5)$$

The first Brillouin zone is illustrated in Figure 2, and it has a hexagonal shape.

## 2.2 Tight Binding Approximation

Each carbon atom in graphene possesses a total of six electrons, including four valence electrons located in the $2s$ and $2p$ orbitals. The $2s$, $2p_x$, and $2p_y$ orbitals of each carbon atom hybridize to form three $sp^2$ orbitals in a trigonal planar arrangement, oriented at 120 degrees to each other. The $2p_z$ orbitals remain unchanged and perpendicular to the plane.

Figure 2: The first Brillouin zone of graphene,

After hybridization, the significant overlap of $sp^2$ orbitals between neighboring atoms leads to the creation of three pairs of $\sigma/\sigma^*$ bands, representing bonding and antibonding covalent bonds, respectively. These bands, distant from the Fermi level, contribute to the stability and strength of graphene in the plane.

In contrast, the less significant overlap between the $2p_z$ orbitals gives rise to the $\pi/\pi^*$ bands, which intersect at what is called the Dirac point. These $\pi$ and $\pi^*$ bands determine the electronic properties of graphene, as they form a collection of semi-empty orbitals both above and below the honeycomb structure. We will, thus, focus on them to determine the dispersion relation and deduce the electronic properties of graphene (Figure 4).

We focus solely on the $p_z$-type orbitals located on all A and B atoms, denoted as $|\Phi_A\rangle$ and $|\Phi_B\rangle$, respectively. The wave functions for these orbitals are expressed as:

$$\langle \mathbf{r}|\phi_A\rangle = \phi_A(\mathbf{r} - \mathbf{R}_j^A),$$

$$\langle \mathbf{r}|\phi_B\rangle = \phi_B(r - \mathbf{R}_j^B).$$

Here, $\phi_A$ and $\phi_B$ represent the $p_z$-orbitals, $\mathbf{r}$ is the position in space, and $\mathbf{R}_j^A$ and $\mathbf{R}_j^B$ are the positions of atoms A and B, respectively.

We describe graphene as the sum of two sublattices of atoms A and B. The eigenstate of the system can be expressed as the sum of two Bloch waves corresponding to the two sublattices A and B.

So the wave function $\Psi^{\mathbf{k}}(\mathbf{r})$ can be expressed as:

$$\Psi^{\mathbf{k}}(\mathbf{r}) = \frac{1}{\sqrt{N}} \sum_j e^{i\mathbf{k}\cdot\mathbf{R}_j} \left(c_A(\mathbf{k})\phi(\mathbf{r}-\mathbf{R}_j^A) + c_B(\mathbf{k})\phi(\mathbf{r}-\mathbf{R}_j^B)\right),$$

which can be further written as:

$$\Psi^{\mathbf{k}}(\mathbf{r}) = c_A(\mathbf{k})\Phi_A^{\mathbf{k}}(\mathbf{r}) + c_B(\mathbf{k})\Phi_B^{\mathbf{k}}(\mathbf{k}),$$

where

$$\Phi_{A/B}^{\mathbf{k}}(\mathbf{r}) = \frac{1}{\sqrt{N}} \sum_j e^{i\mathbf{k}\cdot\mathbf{R}_j} \phi(\mathbf{r}-\mathbf{R}_j^{A/B}).$$

We describe the system's Hamiltonian using the tight-binding approximation method. Each atom is assigned a site energy, denoted $\epsilon_A$ for type-A atoms and $\epsilon_B$ for type-B atoms. The Hamiltonian also includes nearest-neighbor hopping energies, denoted as $\gamma$, between atoms (each atom having three nearest neighbors). The expression for the Hamiltonian is given by:

$$\begin{aligned}H = &\sum_i \left(\epsilon_A |\Phi_A^i\rangle\langle\Phi_A^i|\right) \\ &+ \sum_j \left(\epsilon_B |\Phi_B^j\rangle\langle\Phi_B^j|\right) \\ &- \gamma \sum_{[i,j]} \left(|\Phi_A^i\rangle\langle\Phi_B^j| + |\Phi_B^j\rangle\langle\Phi_A^i|\right)\end{aligned} \quad (2.6)$$

Here, the notation [i, j] indicates that the sum is over nearest neighbors.

By solving the eigenvalue equation:

$$H|\Psi^{\mathbf{k}}\rangle = E(\mathbf{k})|\Psi_{\mathbf{k}}\rangle,$$

we obtain the eigenstates $|\Psi^{\mathbf{k}}\rangle$ and eigenvalues $E(\mathbf{k})$.

$$\begin{aligned}H|\Psi^{\mathbf{k}}(\mathbf{r})\rangle = \sqrt{N} \sum_j e^{i\mathbf{k}\cdot\mathbf{R}_j} &\bigg[\epsilon_A c_A(\mathbf{k})|\Phi_A^j\rangle + \epsilon_B c_B(\mathbf{k})|\Phi_B^j\rangle \\ &-\gamma\bigg((e^{i\mathbf{k}\cdot\boldsymbol{\delta_1}} + e^{i\mathbf{k}\cdot\boldsymbol{\delta_2}} + e^{i\mathbf{k}\cdot\boldsymbol{\delta_3}})c_B(\mathbf{k})|\Phi_A^j\rangle \\ &+ (e^{-i\mathbf{k}\cdot\boldsymbol{\delta_1}} + e^{-i\mathbf{k}\cdot\boldsymbol{\delta_2}} + e^{-i\mathbf{k}\cdot\boldsymbol{\delta_3}})c_A(\mathbf{k})|\Phi_B^j\rangle\bigg)\bigg]\end{aligned} \quad (2.7)$$

Multiplying by $\langle\Phi_{A/B}^j|$, we obtain the system of equations:

$$E(\mathbf{k})c_A(\mathbf{k}) = \epsilon_A c_A(\mathbf{k}) - \gamma(e^{i\mathbf{k}\cdot\boldsymbol{\delta_1}} + e^{i\mathbf{k}\cdot\boldsymbol{\delta_2}} + e^{i\mathbf{k}\cdot\boldsymbol{\delta_3}})c_B(\mathbf{k})$$
$$E(\mathbf{k})c_B(\mathbf{k}) = \epsilon_B c_B(\mathbf{k}) - \gamma(e^{-i\mathbf{k}\cdot\boldsymbol{\delta_1}} + e^{-i\mathbf{k}\cdot\boldsymbol{\delta_2}} + e^{-i\mathbf{k}\cdot\boldsymbol{\delta_3}})c_A(\mathbf{k})$$

We define $f(\mathbf{k}) = -\gamma(e^{i\mathbf{k}\cdot\boldsymbol{\delta_1}} + e^{i\mathbf{k}\cdot\boldsymbol{\delta_2}} + e^{i\mathbf{k}\cdot\boldsymbol{\delta_3}})$

Moreover, we consider the graphene to be completely symmetric, so the energies on sites A and B are equal. We set them equal to zero for simplicity: $\epsilon_A = \epsilon_B = 0$. Thus, the Hamiltonian becomes:

$$H = \begin{pmatrix} 0 & f(\mathbf{k}) \\ f^*(\mathbf{k}) & 0 \end{pmatrix}. \quad (2.8)$$

We obtain the dispersion relation by solving the secular equation $\det(H-E(\mathbf{k})I)=0$, where $I$ is the unit matrix.

This leads to:

$$\det\begin{pmatrix} -E & f(\mathbf{k}) \\ f^*(\mathbf{k}) & -E \end{pmatrix} = 0$$

$$E = \pm f^*(\mathbf{k})f(\mathbf{k}) = |f(\mathbf{k})|$$

With

$$f^*(\mathbf{k})f(\mathbf{k}) = \gamma^2\left(3 + 2\cos\sqrt{3}k_x a + 4\cos\frac{\sqrt{3}}{2}k_x a \cos\frac{3}{2}k_y a\right)$$

Hence, the dispersion relation is:

$$E_\pm(\mathbf{k}) = \pm\gamma\sqrt{3 + 2\cos(\sqrt{3}k_x a) + 4\cos\left(\frac{\sqrt{3}}{2}k_x a\right)\cos\left(\frac{3}{2}k_y a\right)}$$

A graphical representation of this dispersion relation is provided by Figure 3

## 2.3 Massless Electrons: Architects of Graphene's Exceptional Conductivity

We expand the dispersion relation $E_\pm(\mathbf{k})$ to first order around the K-point (or K' or, more generally, around the Dirac points):

$$\begin{aligned} E_\pm(\mathbf{k}) &\approx \pm\gamma\frac{3a}{2}\sqrt{(\delta k_x)^2 + (\delta k_y)^2} + \mathcal{O}(3) \\ E_\pm(\mathbf{k}) &\approx \pm\gamma\frac{3a}{2}|\delta\mathbf{k}| \end{aligned} \quad (2.9)$$

Therefore, we observe that the energy linearly depends on $\delta_k$ around the Fermi level and in the vicinity of the Dirac point.
Denoting $v_F = \frac{3a\gamma}{2\hbar}$ as the Fermi velocity, we obtain :

$$E_\pm = \pm v_F \hbar |\delta\mathbf{k}| \quad (2.10)$$

where the value $v_F = 10^6$ m/s $= \frac{C}{300}$ and $\hbar$ is the reduced Planck constant.

This dispersion is known as the Dirac cone (see the inset in the Figure 3), and it is from here that the exceptional electronic properties of graphene emanate.

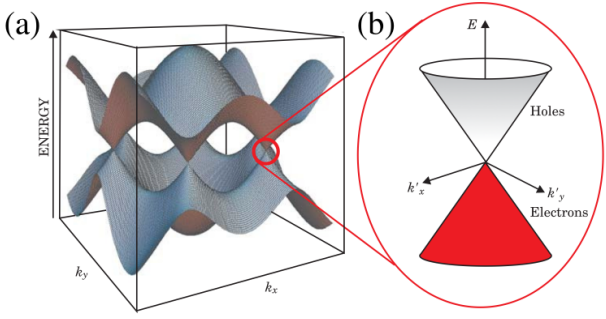

Figure 3: 3D Representation of the $\pi$ and $\pi^*$ Band Structure. Zoom performed at the K point of the first Brillouin zone (Dirac point): Linear dispersion relation known as Dirac cones is observed.

Indeed, the major difference between this result and the usual dispersion relation $E(\mathbf{k}) = \frac{(\hbar \mathbf{k})^2}{2m}$, where $m$ is the electron mass, is that the Fermi velocity does not depend on the electron mass.

The linearity of the dispersion relation is characteristic of relativistic particles such as photons. In the case of electrons in graphene, their velocity is 300 times lower than the speed of light. However, this relation bears a formal analogy to the Dirac equation, in the case of fermions with zero mass. This is why we liken the charge carriers in graphene to Dirac fermions with zero mass.

These effectively massless electrons are highly mobile, and their mobility significantly exceeds that of electrons in silicon, the foundational material in today's electronics. This fact positions graphene as a prime candidate for high-speed electronics.

## 2.4 Chirality and pseudo-spin

In the Hamiltonian of graphene as written in Eq.(2.8), the term $f(\mathbf{k})$ represents the matrix element describing the coupling between the sublattices A and B.

The existence of two equivalent but independent sublattices, A and B, in graphene gives rise to a new property known in the graphene dynamics, which is known as chirality. This chirality is observed in the independence of the two linear branches in the energy dispersion of graphene, which intersect at the Dirac points. This independence indicates the presence of a quantum number, named pseudospin, analogous to the electron spin but entirely separate from the actual spin. In essence, both sublattices contribute to a unique graphene behavior, introducing a kind of fictitious spin that influences electron behavior similarly to real spin but without a direct connection to it.

At the Dirac point K, $f(\mathbf{K}) = e^0 + e^{i2\pi/3} + e^{-i2\pi/3} = 0$. This indicates no coupling between sublattices A and B at the Dirac points.

Also, since both sublattices are hexagonal Bravais lattices of carbon atoms, they share the same quantum states. This similarity results in a degeneracy point at the Dirac points.

The precise cancellation of the three factors of $f(\mathbf{K})$ describing the coupling between sublattices A and B no longer holds when the wave vector is not exactly equal to that of the $\mathbf{K}$ point.

Let's find the expression for $f(\mathbf{k})$ around the Dirac point $\mathbf{K}$.

Consider a point $\mathbf{K} + \mathbf{q}$ very close to $\mathbf{K}$ with $\mathbf{q} = (q_x, q_y)$. Expanding $f(K + \mathbf{q})$ to first order in $q_x$ and $q_y$, we have:

$$f(\mathbf{K} + \mathbf{q}) \approx \frac{3}{2}\gamma a q_x - i\frac{3}{2}\gamma a q_y$$

$$f(\mathbf{K} + \mathbf{q}) \approx \hbar v_F (q_x - i q_y)$$

Using this approximate expression for the function $f(\mathbf{K})$, the Hamiltonian in the vicinity of the point $\mathbf{K}$ becomes

$$H = v_F \begin{pmatrix} 0 & q_x - iq_y \\ q_x + iq_y & 0 \end{pmatrix} = \hbar v_F \boldsymbol{\sigma} \cdot \mathbf{q}.$$

where $\boldsymbol{\sigma} = (\sigma_x, \sigma_y)$ is the vector of Pauli matrices.

The eigenvalues of the energy and the corresponding eigenvectors of $H$ are given by :

$$E_\pm = \hbar \pm v_F |\mathbf{q}| \text{ and } \Psi_\pm = \frac{1}{\sqrt{2}} \begin{pmatrix} 1 \\ se^{i\theta} \end{pmatrix} e^{i\mathbf{q} \cdot \mathbf{r}/\hbar},$$

where $s = \pm$ denotes the conduction and valence bands, respectively.
Here, $\theta$ is the polar angle of the momentum in the graphene plane, $\mathbf{q} = (q_x, q_y) = |\mathbf{q}|(\cos\theta, \sin\theta)$.

## 3 Opening of a Band Gap through Electrical and Magnetic fields

Within the realm of graphene electronics, addressing the electronic limitations of undoped graphene becomes paramount. Undoped ("normal") graphene exhibits the absence of a band gap, presenting challenges and limitations for various applications. In this section, we explore the strategies employed in opening a band gap through electromagnetic modulation, along with the potential implications and applications of such advancements

## 3.1 Electronic Limitations of Undoped Graphene

As mentioned above, graphene does not have a band gap, this means there is an absence of an energy gap between the valence and conduction bands. This lack of a band gap in pure graphene imparts unique properties to the material, but it also represents a significant limitation in the context of electronic applications. The absence of an energy gap makes it challenging to modulate electrical conductivity, thereby limiting the possibilities for creating electronic devices such as transistors, which rely on the ability to control the flow of electrons. This limitation also hinders the design of devices like electronic switches, which require the ability to turn the current on and off. This motivates the need to find effective ways to create a band gap, and the method we will present in the following sections of this document involves applying a combination of electric and magnetic fields.

## 3.2 Band Gap opening in single-layer graphene under the modulation of combined 1D electric and magnetic fields

In this section, we explore, using perturbation theory, a criterion for opening a band gap in graphene through the combined application of electric and magnetic fields. Our study builds upon prior research conducted by Lin et al. [6], where they investigated the opening of a band gap in graphene at the Dirac point under the influence of periodic scalar and vector potentials. Based on symmetry analysis, this study suggests that the double degeneracy at the Dirac point can be lifted when the potentials break both chiral symmetry and time-reversal symmetry. The band gap equation at the Dirac point is obtained analytically using perturbation theory, and we provide it in this document. This band gap equation could be considered as a criterion for the opening of the band gap at the Dirac point.

To initiate our study, we make the following approximations:

- The spatial period of the external potential is much larger than the graphene lattice constant ($\sim 1.42\,\text{Å}$), allowing us to neglect the intervalley scattering.

- We adopt the long-wavelength approximation.

- External potentials (electric and magnetic) are considered weak.

We use $\hbar = c = 1$ to simplify the expressions.
Before applying the combined field, the Dirac-like Hamiltonian is written as (at the Dirac points): $H = v_F \sigma \cdot q$ .

When applying a combined electric and magnetic field to graphene, the terms constituting the new Dirac-like Hamiltonian at the Dirac point $K$ are as follows:

- $v_F(-i\nabla)$: Kinetic term associated with the Fermi velocity.

- $eA$: Coupling term with the magnetic field, where $e$ is the electron charge, and $A$ is the magnetic vector potential. This term represents the interaction between electrons and the magnetic field.

- $eU$: Electric potential term, where $U$ is the electric potential. It represents the interaction of electrons with the electric potential.

Thus, the Dirac-like Hamiltonian around the $K$ point is expressed as follows:

$$H = v_F(-i\nabla + eA) \cdot \sigma + eU \cdot I,$$

where $I$ is the $2 \times 2$ unit matrix.

We choose the Landau gauge such that $A(x) = (0, \int B(x)\,dx, 0)$. Consequently, the Dirac-like Hamiltonian is reformulated as:

$$H = H_0 + H',$$

with

$$H_0 = v_F \begin{pmatrix} 0 & k_x - ik_y \\ k_x + ik_y & 0 \end{pmatrix},$$

and

$$H' = \begin{pmatrix} -eU & -iv_F eA \\ iv_F eA & eU \end{pmatrix}.$$

$H'$ will be treated here as a small perturbation caused by the (weak) electric and magnetic potentials applied to graphene. In the subsequent analysis, we employ stationary perturbation theory at the Dirac point $K$. For simplicity, we assume $\langle B(x) \rangle = 0$, $\langle U(x) \rangle = 0$, and $\langle A(x) \rangle = 0$, ensuring the nullity of the first-order perturbation. Thus, we proceed to consider second-order perturbation effects.
The secular equation for second-order perturbation at the Dirac point is given by:

$$\begin{bmatrix} H_{11} & H_{12} \\ H_{21} & H_{22} \end{bmatrix} \psi - E * I \psi = 0,$$

where

$$H_{ij} = \sum_s \sum_k \frac{\langle \psi_i | H' | \psi_{s,k}^0 \rangle \langle \psi_{s,k}^0 | H' | \psi_j \rangle}{0 - E},$$

whith $\psi_{s,k}^0$ is the eigenstate of $H_0$ (found in section 2.3), and $\psi_i$ is the zeroth-order wavefunction ($i = 1, 2$) at the double-degenerate Dirac point.

The box normalized wavefunctions are given by:

$$\psi_{s,k}^0(x, y) = \frac{1}{\sqrt{L}} \frac{1}{\sqrt{2}} \begin{bmatrix} 1 \\ s e^{i\theta_k} \end{bmatrix} e^{ik_x x} e^{ik_y y},$$

Where $\frac{1}{\sqrt{L}}$ is a box normalization factor, and $\theta_k$ is the polar angle of the wave vector $\mathbf{k} = (|k_x|, |k_y|)$.

Thanks to the previous equations, we may deduce the matrix elements of the secular equation are: $H_{12} = H_{21} = 0$, $H_{11} = -H_{22}$.
We will provide the determination of $H_{11}$ here.
$H_{12}$, $H_{21}$ and $H_{22}$ have been obtained in the same manner

We get :

$$H_{11} = \sum_s \sum_k \frac{\langle \psi_1 | H' | \psi_{s,k}^0 \rangle \langle \psi_{s,k}^0 | H' | \psi_1 \rangle}{0 - E},$$

The summation $\sum_s$ implies that $H_{11}$ should be summed over the entire band $s = \pm 1$ (up and down bands).

For convenience, we divide the entire band into eight symmetric parts based on the signs of energy ($E$) as well as the components of wave vectors ($k_x$ and $k_y$) as follows:
$E > 0$, $k_x > 0$, $k_y > 0$:

$$\psi_1^0 = \frac{1}{\sqrt{L}} \frac{1}{\sqrt{2}} \begin{bmatrix} 1 \\ e^{i\theta_k} \end{bmatrix} e^{ik_x x} e^{ik_y y},$$

$E > 0$, $k_x < 0$, $k_y > 0$:

$$\psi_2^0 = \frac{1}{\sqrt{L}} \frac{1}{\sqrt{2}} \begin{bmatrix} 1 \\ -e^{-i\theta_k} \end{bmatrix} e^{-ik_x x} e^{ik_y y},$$

$E > 0$, $k_x > 0$, $k_y < 0$:

$$\psi_3^0 = \frac{1}{\sqrt{L}} \frac{1}{\sqrt{2}} \begin{bmatrix} 1 \\ e^{i\theta_k} \end{bmatrix} e^{ik_x x} e^{-ik_y y},$$

$E > 0$, $k_x < 0$, $k_y < 0$:

$$\psi_4^0(x,y) = \frac{1}{\sqrt{L}} \frac{1}{\sqrt{2}} \begin{bmatrix} 1 \\ -e^{i\theta_k} \end{bmatrix} e^{-ik_x x} e^{-ik_y y},$$

$E < 0$, $k_x > 0$, $k_y > 0$:

$$\psi_5^0 = \frac{1}{\sqrt{L}} \frac{1}{\sqrt{2}} \begin{bmatrix} 1 \\ -e^{i\theta_k} \end{bmatrix} e^{ik_x x} e^{ik_y y},$$

$E < 0$, $k_x < 0$, $k_y > 0$:

$$\psi_6^0 = \frac{1}{\sqrt{L}} \frac{1}{\sqrt{2}} \begin{bmatrix} 1 \\ e^{-i\theta_k} \end{bmatrix} e^{-ik_x x} e^{ik_y y},$$

$E < 0$, $k_x > 0$, $k_y < 0$:

$$\psi_7^0 = \frac{1}{\sqrt{L}} \frac{1}{\sqrt{2}} \begin{bmatrix} 1 \\ -e^{-i\theta_k} \end{bmatrix} e^{ik_x x} e^{-ik_y y},$$

$E < 0$, $k_x < 0$, $k_y < 0$:

$$\psi_8^0 = \frac{1}{\sqrt{L}} \frac{1}{\sqrt{2}} \begin{bmatrix} 1 \\ e^{i\theta_k} \end{bmatrix} e^{-ik_x x} e^{-ik_y y},$$

The component $H_{11}$ can be expressed as follows:

$$H_{11} = \sum_{s=1, k_x > 0, k_y > 0} \sum_{i=1}^{8} S_i \quad (3.1)$$

Where

$$S_i = \frac{\langle \psi_i^0 | H' | \psi_{s,k}^0 \rangle \langle \psi_{s,k}^0 | H' | \psi_i^0 \rangle}{0 - E},$$

By summing up $\sum_{i=1}^{8} S_i$, we obtain :

$H_{11} = \sum_{s=1, k_x > 0, k_y > 0} I_{11}$ with

$$I_{11} = \frac{1}{L^2} \frac{1}{E} \frac{8 v_F}{\cos \theta_k} \left( \int_0^L dx \, eU(x) \sin(k_x x) \right.$$

$$\int_0^L dx \, eA(x) \cos(k_x x) - \int_0^L dx \, eU(x) \cos(k_x x)$$

$$\left. \int_0^L dx \, eA(x) \sin(k_x x) \right) \quad (3.2)$$

where $k_x = k \cos \theta_k$, $k_y = k \sin \theta_k$, and $k = \frac{E}{v_F}$.

We can then convert the sum $\sum_{s=1, k_x > 0, k_y > 0}$ into an integral, i.e.,

$$H_{11} = \int_{s=1, k_x > 0, k_y > 0} dE \, d\theta_k \, \rho(E, \theta_k) \, I_{11},$$

where $\rho(E, \theta_k) = \frac{2 A_c E}{\pi} \cdot \frac{1}{2\pi k} = \frac{A_c}{\pi^2 v_F}$,
and $A_c$ represents the area of the honeycomb lattice's unit cell in graphene.

The eigenvalues of the secular equation, easily obtained from equation (A7), are given by $E = \pm H_{11}$.

It is clear that when $H_{11} \neq 0$, the double degeneracy at the Dirac point is broken, and a band gap is generated at the Dirac point. Thus, equation (3.1) can be considered as a criterion for the opening of the band gap at the Dirac point.

The electronic properties of graphene are affected by a variety of phenomena, let us explore some examples in the following sections.

## 4 Spin-orbit Coupling and Spin-Hall Effect

Spin-orbit coupling describes a process in which an electron spontaneously changes spin and angular momentum, or switches from one orbital wave function to another. The mixing between orbital motion and spin is a relativistic effect, which is extracted from the Dirac model for electrons. The effect is predominant in heavy ions, as the speed of the electrons is generally higher. But for carbon, which is a light atom, the spin-orbit coupling is weak. The Dirac point is where a gap is formed, allowed by the symmetries of spin-orbit interaction. The appearance of this gap leads to non-trivial spin hall conductance. This leads to a spin-dependent shift of the suborbitals, acting as an effective mass inside each Dirac point.

The spin Hall effect is a transport phenomenon consisting of the appearance of spin accumulation on the lateral surfaces of a sample carrying electric current. When the inversion symmetry (invariance under a point reflection) of the honeycomb lattice is broken, either because the graphene layer is curved or because an external electric field is applied, additional terms that modulate the nearest neighbor's jump are induced.

The magnitude of the spin-orbit coupling in graphene can be deduced from the known spin-orbit coupling in the carbon atom. This coupling allows transitions between the $p_x$, $p_y$, $p_z$ orbitals. ($p_x$ orbital has two lobes oriented along the x-axis and $p_y$ orbital has two lobes oriented along the y-axis, whereas $p_z$ orbital has two lobes oriented along the z-axis).

An electric field also induces transitions between $p_z$ and $p_x$ orbitals. These intra-atomic processes mix the $\pi$ and $\sigma$ bonds in graphene. Sigma bonds result from an overlap of sp2 hybrid orbitals (Orbital overlay, covalent bond) whereas $\pi$ bonds emerge from tunneling between the protruding $p_z$ orbitals (see Figure 4, for clarity, only one $p_z$ orbital is shown with its three nearest neighbors)

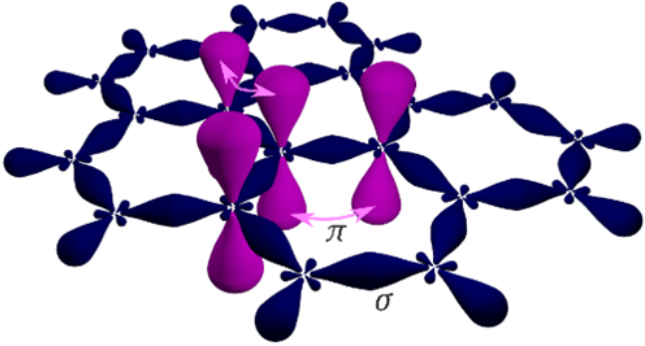

Figure 4: Orbit structure of graphene

## 5 Disorder in Graphene

Second disorder practically affects graphene in the following way: A deformation in the carbon grid causes a change in the cell unit, due to expansion or contraction. This change in electron density induces a local change in the chemical potential. One of the properties that makes Graphene electronically special is the robustness of sigma bonding. It is physically and thermodynamically difficult to insert something into the lattice. But it is always possible for imperfections to take place, which is the basis of extrinsic and intrinsic properties.

### 5.0.1 Intrinsic Sources

- Surface undulations (temperature gradient, interaction with the environment, substrate, scaffolding and absorbents. In all cases, the disorder is induced by a change in the distance and the interatomic relative angle. Dirac fermions propagate in waves on the Graphene by a potential proportional to the square of the curvature.This is why Graphene has increased resistivity.

- Using scanning tunneling microscope (STM), topological defects violate of crystal symmetry.

### 5.1 Extrinsic Sources

Adatomes: adsorbed extra atoms, vacant site, extensive defects such as cracks and edges. Imperfections and their effects are not negligible in Graphene to explain the modifications of a signal propagating on it. All in all, defects cause a change in the wave functions and propagation of electrons. More info: Guinea et al.,2005

## 6 Conclusion

We have given a quick oversight of graphene's exceptional electronic properties. Even though what we presented is but a fraction of what is known, it should have been enough to convince the reader of the potential this material possesses. As we mentioned earlier, graphene is also remarkable when considering its mechanical resistance, the superconductivity effects that have been observed, and its thermal properties. A clear sign that it promises to be exceptionally important in the future is the EU's 1 billion dollar Graphene Flagship, which aims to explore further into some of the effects we presented, among many others.

Regarding the electronic properties of graphene, some points presented deserve to be highlighted: the massless electrons and the possibility of modifying graphene to correct the eventual shortcomings it may have. These make the strength of the material and allow us to be optimistic for the future it may bring.